\documentclass[twocolumn,astrosymb,usenatbib]{aastex702}
\usepackage{xcolor}
\usepackage{amsmath}	
\usepackage{amssymb}	
\usepackage{physics}   
\usepackage{verbatim}
\usepackage{subcaption}

\usepackage{float}

\begin{document}

\title[Dynamical ages for young stellar systems]{Investigating the timescales traced by dynamical ages in young stellar systems}

\author[0009-0006-1587-719X]{Isabella Armstrong}
\affiliation{Department of Physics and Astronomy, McMaster University, 1280 Main Street West, Hamilton, ON, L8S 4M1, Canada}
\email[show]{armstri@mcmaster.ca}

\author[0000-0003-3551-5090]{Alison Sills}
\affiliation{Department of Physics and Astronomy, McMaster University, 1280 Main Street West, Hamilton, ON, L8S 4M1, Canada}
\email{asills@mcmaster.ca}

\author[0000-0002-6116-1014]{Claude Cournoyer-Cloutier}
\affiliation{Max Planck Institute for Astrophysics, Karl-Schwarzschild-Stra\ss e 1, 85748 Garching bei M\"unchen, Germany}
\email{cournoyc@MPA-Garching.MPG.DE}

\begin{abstract}

Stars form in clustered environments, and then the stars may congregate into star clusters. The timescale on which individual stars form is not necessarily the same as the timescale on which star clusters assemble. Therefore, developing methods to probe these different timescales is a crucial part of understanding star cluster assembly and early evolution. Dynamical ages can be constructed from the kinematics of individual stars in clusters in the solar neighborhood using Gaia data. Using a star formation simulation, we test five different dynamical age methods. We divide dynamical ages into two approaches: methods which trace the size evolution of the cluster, and methods which trace the trajectories of individual stars. We confirm that size methods cannot accurately derive ages in bound clusters whereas methods tracing individual stars can be applied to both bound and unbound systems. We find that different dynamical age methods trace different epochs of star cluster formation and assembly: The method using stars with the oldest traceback ages provides an upper limit on stellar ages and the method using the median traceback ages is the best proxy for isochronal ages. Methods which trace the size evolution of the cluster best trace the onset of cluster expansion. Our work suggests that the median traceback time of stars moving away from the cluster provides the most reliable proxy for isochronal age, and that this method can be applied to any expanding system. 

\end{abstract}


\section{Introduction}\label{sec:intro}

Most stars form in clustered environments, from turbulent clouds of dense molecular gas \citep{Lada&Lada2003, Krumholz2019}. Star clusters and associations, seen observationally as local over-densities of stars may then assemble out of the structure in the collapsing cloud prior to the expulsion of the natal gas by feedback \citep{Kuhn2019, Guszejnov2022, Claude2023, arnold2024}. These young clusters may be gravitationally bound or unbound, and may begin bound before becoming unbound at a later stage through dynamical interactions or changes in the local potential \citep{Krumholz2019}. 

Understanding the timescales on which star clusters assemble is key to understanding the population of star clusters we observe in the solar neighborhood. It is often assumed that star clusters form as simple stellar populations with a single age \citep{Bell2015, Beasor2019}. Techniques to derive ages for observed clusters are usually based on properties internal to each star, and include isochrone fitting \citep[e.g.][]{Baraffe2002,Bell2015,Luhman2022}, lithium depletion boundary \citep[e.g.][]{Mentuch2008,Tognelli2015,GalindoGuili2022}, gyrochronology \citep[e.g.][]{Bouma2024,Lu2024}, and asteroseismology \citep[e.g.][]{Miglio2021}. However, these methods can suffer from uncertainties inherent to model parameters \citep[e.g. uncertain composition of the stars,][]{Bell2015}, the adopted input physics \citep[e.g. nuclear reaction cross-sections,][]{Tognelli2015}, or systematic uncertainties in observational constraints \citep{Miglio2021}. The most common method for assigning an age to clusters is isochrone fitting. However, isochronal ages are especially uncertain for young clusters ($< 50$ Myr) \citep{Guzman2026}. In particular, observations of young and intermediate age clusters have shown features like multiple main sequences and stars more luminous than the nominal turnoff, suggestive of limitations in the simple stellar population models \citep{Li2017}. Simple stellar population models do not account for  binary evolution and interactions, which may be responsible for these discrepancies \citep{Wang2020binaries}. Furthermore, \cite{Beasor2019} showed that fitting isochrones to different features in colour-magnitude diagrams can yield different inferred ages. Disagreement between different age determination methods has driven the search for additional tools to describe cluster ages \citep{Kerr2025}. 

Recently, the European Space Agency's \textit{Gaia} mission \citep{GaiaDR3} has made it possible to undertake detailed kinematic studies of nearby groups and clusters \citep{Miret-Roig2020, Kerr2023, Couture2023, Pelkonen2024, armstrong2025}. \textit{Gaia} provides 6D kinematics for millions of stars, although radial velocity measurements are often supplemented with observations from other telescopes due to the higher uncertainties in Gaia radial velocities compared to those that can be obtained from the ground \citep[e.g.][]{Kerr2022, armstrong2025}.

Young stellar systems are expected to preserve the kinematic properties of the stars when they formed. Therefore, precise 6D kinematic information may encode valuable information about the cluster assembly processes. Recent works use stellar kinematics to perform dynamical traceback analysis, tracing the trajectories of the stars backwards in time to derive dynamical ages. Dynamical ages are independent of stellar evolutionary models. Instead, these techniques use the trajectories of either individual stars or all cluster members simultaneously, utilizing the assumption that the stars formed at the same time and in the same place, and that stellar velocities are constant in time. \citep[see][]{Miret-Roig2020, Quintana2022, Couture2023, Kerr2023, Pelkonen2024, Galli2024, Miret-Roig2024, armstrong2025, Cheshire2026}.  

Dynamical ages are often compared with other age metrics, such as isochrone fitting, to better constrain when a cluster formed. However, at present there is little consensus in the literature about what constitutes a dynamical age. Different works use different methods making it unclear what physical processes are measured by each unique dynamical age method. Techniques may be designed to trace the isochronal evolution of the star clusters, with the goal of tracing when the stars formed \citep{Pelkonen2024}. Other works highlight the differences between the isochronal ages and the kinematic traceback timescales and suggest that this discrepancy may probe the timescale on which physical processes such as gas expulsion may occur \citep{Miret-Roig2024}. \par 

Observations provide a single snapshot of a system at the present time. In contrast, detailed star cluster formation simulations can trace the evolutionary history of star clusters from the natal gas cloud to gas expulsion  \citep[e.g.][]{Grudic2021, Claude2024}. In this work, we apply five different dynamical age methods to a simulation of star clusters formed under conditions typical of the disk of the Milky Way \citep{Claude2024}. Our goal is to understand how dynamical ages compare to each other, to the ages of the stars within the cluster, and to different stages of cluster assembly. We compute dynamical ages for clusters formed in a simulation where the formation and evolution of the cluster and its stars is known. Using a simulation allows us to test how dynamical ages evolve over time and investigate which dynamical age methods can quantify the signatures of star cluster formation and assembly in a meaningful way.

This paper is structured as follows: section \ref{sec:methods} describes the simulation and the different dynamical ages investigated in this work. Section \ref{sec:results} presents our results, and their significance is discussed in section \ref{sec:discussion}. 

\section{Methodology} \label{sec:methods}

\subsection{Star cluster formation simulation}\label{sec:methods-sim}

We use a cloud-scale cluster formation simulation presented in \cite{Claude2024} and briefly describe it here. The simulation was performed using \textsc{Torch} \citep{Wall2019, Wall2020} which couples MHD to star formation, stellar dynamics, stellar evolution, heating and cooling, and stellar feedback (UV radiation pressure, ionization, and stellar winds) through the \textsc{Amuse} framework \citep{Portegies2009, Pelupessy2013, Portegies2018}. MHD is computed using FLASH \citep{Fryxell2000, Dubey2013} and collisional stellar dynamics is handled using \textsc{PeTar} \citep{Wang2020}. Star formation is modeled by spawning stars from sink particles as described by \cite{Wall2019}, with the addition of primordial binary formation described in \cite{Claude2021}. 

The cloud has an initial gas mass of $2\times10^4M_\odot$ and an initial radius of $R = 7$ pc. The simulation is evolved in \textsc{Torch} for 4.45 Myr as described in \cite{Claude2024}. At the end of the \textsc{Torch} simulation, the model has reached gas expulsion. 4658 stars have been formed, with a total stellar mass of 7378 $M_\odot$.

We take the simulation from \citet{Claude2024} and further evolve it using the N-body gravitational dynamics code \textsc{PeTar}, which we run through the \textsc{Amuse} framework \citep{Wang2020}. \textsc{PeTar} uses three different N-body algorithms. Long-range interactions are computed using the Barnes-Hut tree method \citep{Barnes1986}, short-range interactions are handled with a fourth-order Hermite integrator \citep{Makino1992}, and close-distance multiple systems are handled using the slow-down algorithmic regularization (SDAR) method \citep{Wang2020SDAR}. The changeover radius for the different regimes is set by the user. The gravitational force between stars more distant than $r_{out}$, set to 0.001 pc, are calculated using the Barnes-Hut tree while forces between stars closer than $r_{in} = 0.1r_{out}$, are calculated using the fourth-order Hermite integrator. Forces between stars with distance between $r_{in}$ and $r_{out}$ are calculated using a combination of the Barnes-Hut tree and Hermite codes. The radius $r_{bin}$ at which the SDAR method is applied is set to $r_{bin} = r_{in}/1.25$. We allow the code to automatically determine the typical tree time step, determined to be 10.5 yr. 

To ensure continuous dynamical evolution between the \textsc{Torch} and the \textsc{PeTar} regimes, we use the positions and velocities of the stars in the 10th last snapshot from \cite{Claude2024} to set our initial conditions in \textsc{PeTar}. We begin the \textsc{PeTar}-only run at $t_{sim} = 4.35$ Myr so that the two regimes overlap by 0.1 Myr. In the final analysis, the \textsc{Torch} output is used until $t_{sim} = 4.45$ Myr. After $t_{sim} = 4.45$ Myr, the \textsc{PeTar} only simulations are used. After 4.35 Myr, no new stars are formed in \textsc{Torch}. Stellar evolution is not included in the \textsc{PeTar} only portion of the simulation as that will not significantly affect the trajectories of the stars over such short (20 Myr) timescales. While we expect fewer than 1\% of our stars (15\% by mass) to explode as a supernova in this time, most of the ones that will explode will do so late in the simulation and therefore will not substantially change the overall dynamical evolution of the remaining stars.

\begin{figure}
    \centering
    \includegraphics[width=0.95\linewidth]{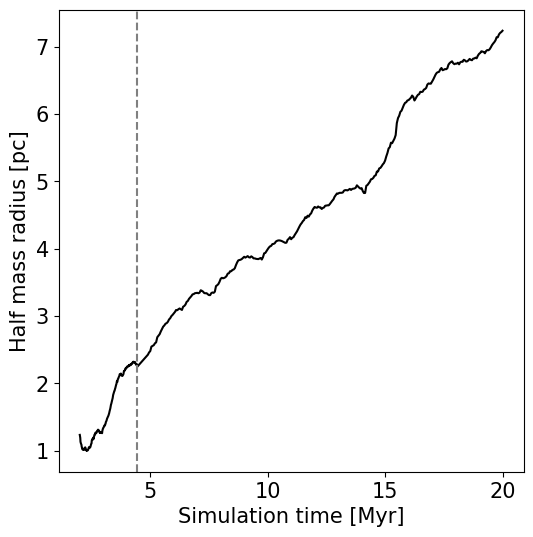}
    \caption{The half mass radius of the entire simulation as a function of simulation time. The vertical gray line plots the \textsc{Torch} to \textsc{PeTar} transition, which occurs at 4.45 Myr.}
    \label{fig:half-mass}
\end{figure}

\begin{figure}
    \centering
    \includegraphics[width=0.95\linewidth]{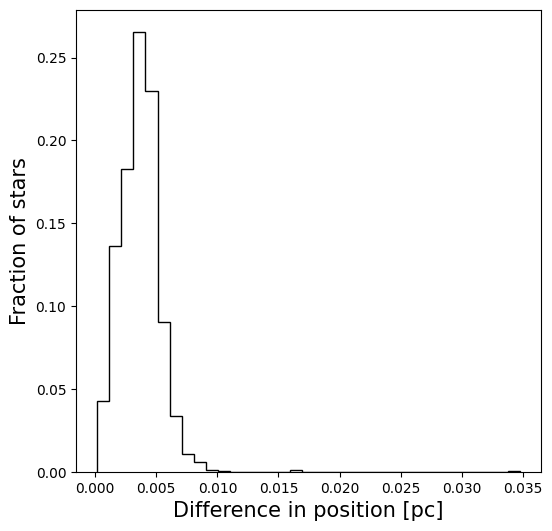}
    \caption{Difference in the positions of the stars in the \textsc{PeTar}-only run compared to the full \textsc{Torch} evolution at $t_{sim} = 4.45$ Myr.  Most of stars have a difference in position of 0.005 pc or less.}
    \label{fig:verify-petar}
\end{figure}

\begin{figure*}
    \centering
    \includegraphics[width=0.99\linewidth]{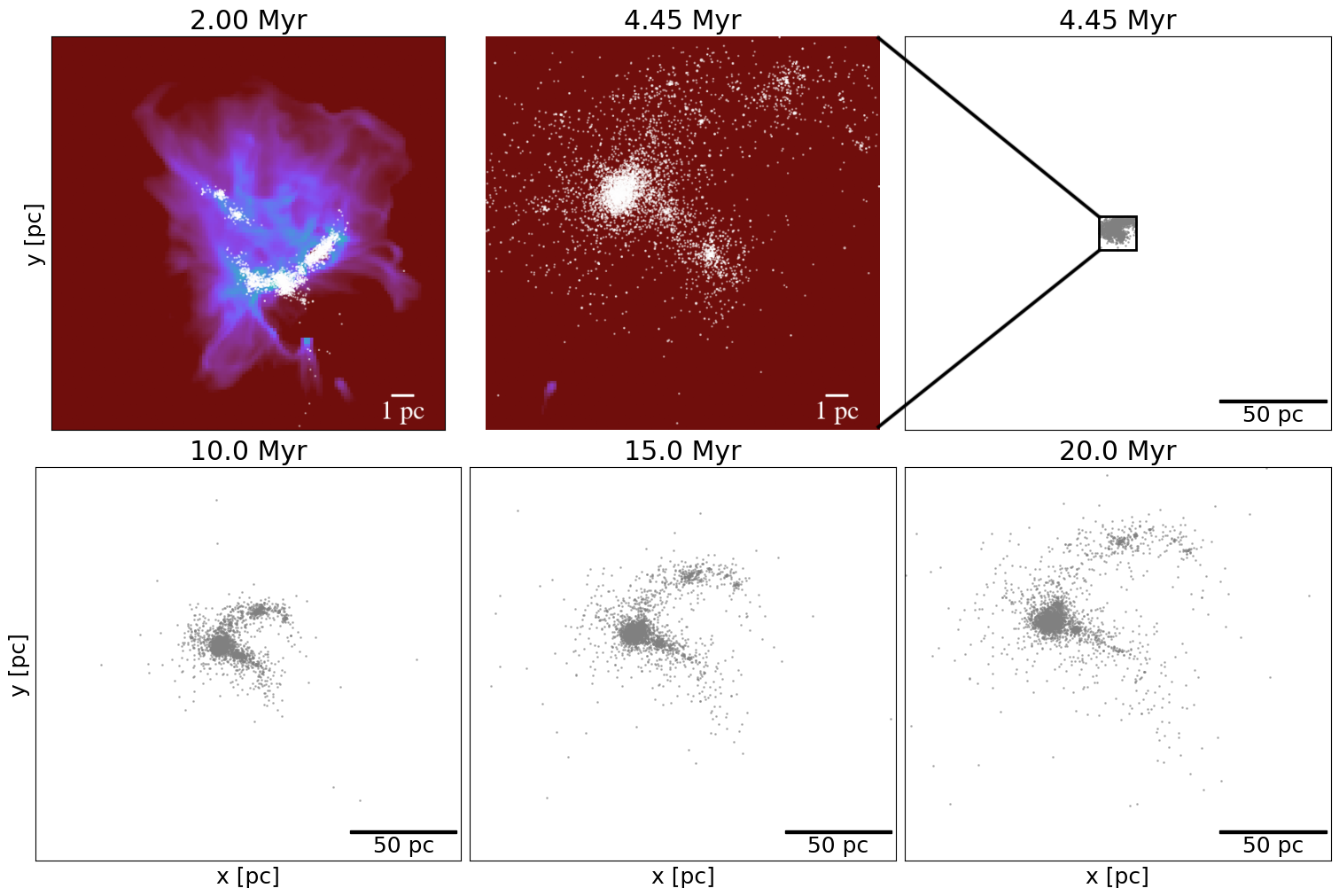}
    \caption{Spatial configuration of the stars within the simulation for six simulation snapshots. At 2 Myr, stars (white dots) are still actively forming. Projected gas density is plotted in the blue colour scale, with brighter regions corresponding to denser gas. At 4.45 Myr, the cluster has reached gas expulsion. After 4.45 Myr, gas was neglected and the simulation was run using the N-body code \textsc{PeTar} only, as show in the last panel of the top row. The box in the right panel of the top row is 17.5 pc on a side, the same size as the two images of the \textsc{Torch} simulations on the first row.}
    \label{fig:time_evolution}
\end{figure*}

The cluster half mass radius, plotted in fig. \ref{fig:half-mass}, shows continuous evolution between the two regimes. In fig. \ref{fig:verify-petar}, we compare the differences in stellar positions in the Torch simulation (with gas) to the PeTar-only simulation at $t_{sim} = 4.45$ Myr. We confirm that neglecting the effects of the small amount of remaining gas beyond 4.35 Myr does not affect the overall evolution of the cluster. We find that the mean difference in position between the two codes is 0.0036 pc, which is smaller than the gas resolution in TORCH (0.13 pc) and the cluster's half mass radius (2.25 pc). Six stars have differences in position greater than 0.01 pc. Two of those stars are in binaries and the remaining four were involved in a stellar merger, after which only one star is retained by \textsc{Torch}. We conclude that the transition between codes is sufficiently smooth, and does not significantly alter the dynamics of the stars within the simulation.  We then evolve the simulation until 20 Myr, comparable in age to the $\beta$ Pictoris moving group, for which dynamical ages have been calculated by \cite{Miret-Roig2020} and \cite{Couture2023}. Fig.~\ref{fig:time_evolution} shows the evolution of the cluster. The first two images in the top row show the stars (white dots) and projected gas density (blue colour scale) from the \textsc{Torch} simulations. The next four panels with grey star show the results from the \textsc{PeTar}-only runs, which cover a much larger spatial extent than the \textsc{Torch} simulation due to expansion of the system.

\subsection{Dynamical ages}\label{sec:methods-dynages}

All dynamical-age based methods make the same assumption that the present day velocity of a star is the same velocity with which it formed. Given a star with present day position $\vec{x}_0$ and present day velocity $\vec{v}_0$, its position some time $t$ ago is given by
\begin{equation}\label{eqn:position}
    \vec{x}(t) = \vec{x}_0 - t\vec{v}_0
\end{equation}

Eqn. \ref{eqn:position} explicitly assumes that there has been no acceleration of the star at any point in the past. In some observational studies, a Galactic potential is considered when reconstructing stellar trajectories. However, \citet{Miret-Roig2020} and \citet{Couture2023} find that the choice of the Galactic potential has a negligible effect on the resultant dynamical age, as the sizes of the clusters are quite small. We test the assumption of constant velocities in Section \ref{sec:results-velocities}.

Dynamical age methods can be broadly divided into two different approaches. We refer to methods which trace the configuration of the cluster back in time using the trajectories of all the member stars  as \textit{size metrics}. This class of methods works by constructing a parameter to determine the size of the cluster at a point in the past (sec. \ref{sec:methods-traceback}). Methods which use the trajectories of individual stars to determine when they were closest to the cluster's center are denoted as \textit{star-by-star methods} (sec. \ref{sec:methods-evap}). 

\subsubsection{Size metric ages}\label{sec:methods-traceback}
Methods which use the size of the cluster assume that the present day cluster is unbound and expanding. Therefore, these methods assume that at some point in the past the stars in the system were closer together than their observed configuration. An age is then determined by tracing the configuration of the stars backwards in time, and finding the point in the past when the size of the cluster was minimized. Size metric methods are expected to trace the time since the cluster became gravitationally unbound, which is not necessarily the time since star formation \citep{Couture2023}.

\begin{flushleft}
    \textbf{Trace and determinant age}
\end{flushleft}

The first dynamical age method we explore is the dynamical traceback age presented in \cite{Miret-Roig2020} and used in subsequent studies \citep[e.g.][]{Couture2023,Galli2024}. A traceback age can only be used on unbound, expanding systems. Traceback ages are also highly sensitive to kinematic outliers \citep{Couture2023}. 

We investigate the two most robust size metrics presented in \cite{Miret-Roig2020}, the Trace Covariance Matrix size $(S_{TCM})$ (eqn. \ref{eqn:trace_size}) and the Determinant Covariance Matrix size $(S_{DCM})$ (eqn. \ref{eqn:det_size}),

\begin{equation}\label{eqn:trace_size}
    S_{TCM} = \bigg[\frac{Tr(\Sigma)}{3} \bigg]^{1/2}
\end{equation}

\begin{equation}\label{eqn:det_size}
    S_{DCM} = \bigg[Det(\Sigma)\bigg]^{1/6}
\end{equation}

where $\Sigma$ is the covariance matrix of the cluster in the spatial coordinates $(x,y,z)$. The corresponding dynamical ages are $\tau_{\text{trace}}$ for the age computed using $S_{TCM}$ and $\tau_{\text{det}}$ for the age computed using $S_{DCM}$. \cite{Miret-Roig2020} used two different estimators of the covariance matrix, the empirical covariance estimation and the robust covariance estimation as implemented in the package \texttt{sklearn.covariance} \citep{Pedregosa2011}. The empirical covariance, \texttt{sklearn.covariance.EmpiricalCovariance}, is the traditional maximum likelihood covariance estimator whereas \texttt{sklearn.covariance.MinCovDet} is the robust estimator which is less sensitive to outliers in the data set as it is designed to estimate to covariance from the subset of data which is tightly distributed. While this approach minimizes the impact of outliers, it becomes computationally expensive for large systems of thousands of stars.

Therefore, in smaller systems we use the robust covariance algorithm and in larger systems of $\geq 1000$ stars we use the empirical covariance algorithm. Because these larger systems consist mainly of bound stars, the choice of covariance estimator does not affect the overall performance of the method, and we include it in our results for comparison to the star-by-star methods. For smaller systems of $\sim 100$ stars, we compare both the empirical and robust covariance estimator. The robust estimator is less sensitive to the effects of individual stars, and we use the robust estimator for our final analysis of these smaller systems.  

To determine when the size estimate of the configuration is minimized, the positions of the stars within the cluster are traced back in time from the present day at $t = 20$ Myr to $t = 0$ Myr in steps of $0.01$ Myr. At each step, the size of the cluster is evaluated. The dynamical age of the system is the value of $t$ at which the size of the system is minimized, that is $\tau_{\text{trace}} = \text{min}(S_{TCM})$ and $\tau_{\text{det}} = \text{min}(S_{DCM})$. 

\begin{flushleft}
    \textbf{Sum of 3D distances}
\end{flushleft}

The next size metric investigated is the sum of the 3D distances between stars, described in \cite{Quintana2022} and used in \cite{armstrong2025}. The positions of the members stars are traced back in time following eqn.~\ref{eqn:position}. At each time step, the distance from each star to the median position of all the stars within the cluster is computed. These distances are summed at each time step. The most compact configuration is the time at which the sum is minimized, and is the resulting dynamical age, $\tau_{\text{dsum}}$. 

\subsubsection{Star-by-star traceback}\label{sec:methods-evap}

The second class of methods we investigate follow the trajectories of individual stars and use the distribution of their individual traceback ages to determine an age for the cluster. First, the position of the cluster center and the velocity of the cluster is defined. The velocity of the cluster is assumed to be constant. Then, the trajectories of individual stars are traced backwards until the time at which the distance between the star and the cluster center is minimized. Each individual star gets its own age, and the cluster is assigned an age based on the distribution of individual stellar traceback ages. The exact definition of the cluster center varies with method, as does the subset of stars used to calculate the age. 

These methods do not require the system to be unbound and expanding. One fundamental assumption of this class of methods is that the overall motion of the cluster's center is also constant in time, i.e. that the relative velocity of each star with respect to the center of the cluster is constant. 

\begin{flushleft}
    \textbf{Evaporation ages}
\end{flushleft}

First presented in \cite{Pelkonen2024}, the evaporation age method makes the assumption that a cluster begins losing (or evaporating) stars immediately after the onset of star formation. The position $(\vec{x}_c)$ and velocity $(\vec{v}_c)$ of the cluster's centre is determined by taking the mean position and velocity of all the stars in the cluster. Next, the median radius $R_{50}$ is defined as the median distance of the stars to the centre of the cluster, $\vec{x}_c$. Then, the trajectory of each individual star outside of $R_{50}$ is traced back in time until the distance between the star and the cluster's centre is minimized. The trajectories of the stars and the motion of the cluster center are reconstructed using eqn. \ref{eqn:position}. Any star whose minimum position is also outside $R_{50}$ is discarded from the rest of the analysis. The remaining stars, called evaporated stars, are those which are presently located outside of $R_{50}$, and whose trajectories trace back within $R_{50}$. A star's evaporation age is the time at which the distance between the star and the cluster’s center is minimized. The 70th percentile of the star evaporation ages, $t_{70}$, is computed and stars with ages greater than $t_{70}$ are kept. The evaporation age $(\tau_{\text{evap}})$ of the cluster is defined as the median evaporation age of these remaining stars. \par 

\begin{flushleft}
    \textbf{Time of closest approach} 
\end{flushleft}

\cite{Cheshire2026} take a similar approach to \cite{Pelkonen2024} by finding the time at which each star most closely approach the cluster center. This method is similar to the two-dimensional star-by-star method used in \citet{armstrong2024}. First, stars which display strong signs of outward expansion are selected. This is determined by computing the angle between the stars' velocities relative to the cluster frame and the vector between the cluster center of mass position and the stars' positions, denoted as $\Delta \theta$. Stars moving radially away from the center of the cluster have $|\Delta \theta| = 0^\circ$ and stars moving radially towards the center of the cluster have $|\Delta \theta| = 180^\circ$. Stars with $|\Delta \theta | < 50^\circ$ are kept. The population is further restricted by selecting stars whose present day position is outside the cluster's half-mass radius, and whose final closest approach distance is within the half-mass radius. For each star, the time of closest approach is when the distance between the star and the cluster's center is minimized. The cluster age, $\tau_{\text{tca}}$, is the median of the closest approach ages. 

\subsection{Clustering}\label{sec:methods-cluster}

\subsubsection{Cluster identification}

\begin{figure}
    \centering
    \includegraphics[width=0.84\linewidth]{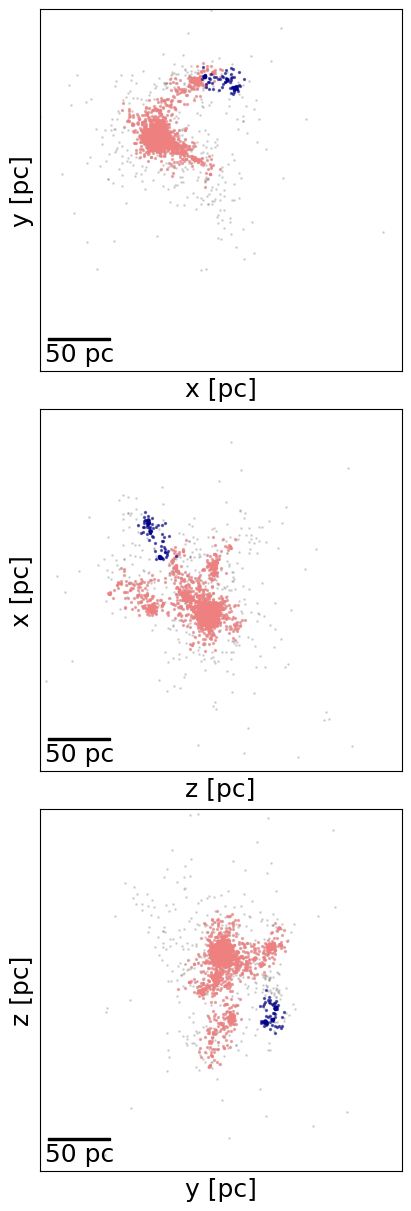}
    \caption{Result of DBSCAN clustering in 3 spatial projections. Pink points show stars belonging to $c_1$, blue points show stars belonging to $c_2$, and grey points show all other stars. The black scale bar marks 50 pc.}
    \label{fig:clustering_20}
\end{figure}

In order to apply these methods to our simulations, we must first identify clusters in the simulation volume. We use the clustering algorithm DBSCAN \citep{Ester1996,Pedregosa2011,Schubert2017}. We adopt the approach taken by \citet{Laverde2025},applying DBSCAN in three-dimensional position (x,y,z) space. Stars are required to have at least five neighbours within a distance $d$, equal to the knee in the distribution of distances to the fifth nearest neighbour, which we determine to be $d = 7.5$ pc for our 20 Myr snapshot. DBSCAN is able to effectively identify clusters of arbitrary shape, without requiring the number of clusters to be specified in advance. Applying DBSCAN to the simulation at 20 Myr identifies one massive cluster and 5 smaller clusters. Fig. \ref{fig:clustering_20} shows the result of the clustering. The most massive cluster contains $\sim 4000$ stars while the second most massive cluster contains $\sim 100$ stars. The most massive cluster, $c_1$, is mostly comprised of bound stars, with a few unbound stars around the edges. In contrast, the second cluster, $c_2$, is mostly unbound. The trend of one massive bound cluster and several unbound clusters holds when varying $d$. We limit our analysis to the two most massive clusters, following stars which end up in $c_1$ and $c_2$ at 20 Myr. This allows for comparison of dynamical ages on both a bound and unbound system. 

\subsubsection{Binary systems}\label{sec:methods-kin_cuts}

Ages which are derived from a size metric are extremely sensitive to contamination of astrometric and kinematic data, which can artificially lower ages \citep{Miret-Roig2020, Kerr2022, Couture2023}. This contamination can come from kinematic outliers, whose kinematic traceback places them far away from the rest of the cluster members, or unresolved multiple systems. 

Our sample includes both primordial and dynamically formed binaries. We replace binaries with semi-major axes less than 10,000 AU with their center of mass positions and velocities, effectively treating the binaries as single stars. We tested removing binaries entirely and found no change in our overall conclusions about the evolution of different dynamical age methods. In observed systems, stars showing signs of binarity (large astrometric deviations from single-source solutions or significant variation in radial velocity between different epochs) are removed from the analysis \citep[typically during the membership selection phase of kinematic studies, see][]{Miret-Roig2020,Kerr2023, Couture2023, Pelkonen2024, armstrong2024, armstrong2025}.

\section{Results}\label{sec:results}

We analyze the two most massive clusters identified with DBSCAN at $t_{sim} = 20$ Myr. Dynamical ages are calculated for each cluster as a function of simulation time. We take the median stellar age of each cluster as a proxy for isochronal age. After $t_{sim} =2.67$ Myr, both clusters show signs of expansion, making them candidates for dynamical age methods. The expansion of the cluster is quantified by performing linear fits to its members in position-velocity space \citep{armstrong2024, dellacroce2024}.

\subsection{Dynamical ages in the simulation}\label{sec:results-age-plots}

\begin{figure*}[h]
    \centering
    \includegraphics[width=0.95\linewidth]{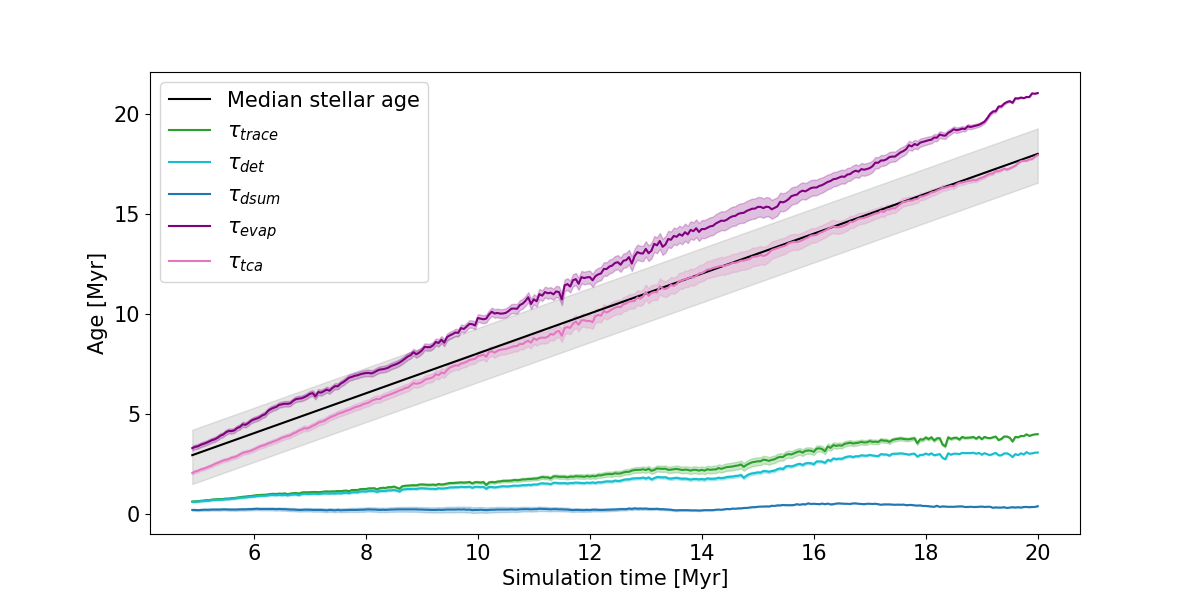}
    \caption{Evolution of different age metrics for $c_1$ as a function of simulation time. Pink and purple lines plot the star-by-star metrics. The blue, green, and teal lines plot the size metric ages. The solid black line plots the median stellar age, with the grey band marking the age of the youngest and oldest star. Typical errors computed using bootstrap error estimation are shown as colored bands around the lines, with a mean of 0.11 Myr and a maximum of 0.5 Myr.}
    \label{fig:ages_c1}
\end{figure*}

The first cluster we analyze, $c_1$, is mostly bound with only 12\% of its stars unbound from the cluster (11\% by mass) and has a median formation time of $t_{sim} = 2.0$ Myr. It contains 4184 stars with a total mass of $6557$ M$_{\odot}$. The first stars formed at $t_{sim} = 0.70$ Myr and the last formed $t_{sim} = 3.42$ Myr. At 20 Myr, $c_1$ has a positive expansion gradient in position-velocity space.  
Fig. \ref{fig:ages_c1} plots dynamical age versus simulation time for $c_1$. For each simulation output, a dynamical age is calculated using the current positions and velocities of the stars flagged as belonging to $c_1$ at 20 Myr. Errors are estimated with the jackknife error estimation method, by omitting each star once and repeating the fit \citep{Hogg2010}. Linear fits are computed for each dynamical age method, with a slope of 1 indicating that the dynamical age is increasing at the same rate as the simulation time and median age. 

The three size methods have slopes of $0.223 \pm 0.001$, $0.162 \pm 0.001$, and $0.014 \pm 0.001$ for $\tau_{\text{trace}}$, $\tau_{\text{det}}$ and $\tau_{\text{dsum}}$ respectively. These slopes are not compatible with a slope of 1, indicating that they do not trace the isochronal age, and fail to estimate the age of $c_1$ correctly, as expected from the bound nature of $c_1$. Despite having a positive expansion gradient, $c_1$ is mainly comprised of bound stars meaning the majority of stars are not moving radially away from $c_1$. Therefore, for any given snapshot, tracing the stars in $c_1$ linearly backwards in time will result in a larger cluster. 

We find that the star-by-star methods have slopes that are closer to 1. $\tau_{\text{evap}}$ has a slope of $1.170\pm 0.001$ and is always greater than $\tau_{\text{tca}}$, which has a slope of $1.043 \pm 0.001$. $\tau_{\text{tca}}$ is the most consistent with the median stellar age of $c_1$, particularly at later times. $\tau_{\text{evap}}$ increases faster than the median stellar age of $c_1$, and can only act as an upper limit on the isochronal age of $c_1$. 

The second cluster, $c_2$, is mostly unbound with a median formation time of $t_{sim} = 1.79$ Myr. It contains 92 stars with a total mass of $177 M_\odot$. The first stars belonging to $c_2$ formed at $t_{sim} = 1.12$ Myr and the last formed at $t_{sim} = 2.07$ Myr. Like $c_1$, $c_2$ also displays a positive expansion gradient making it a suitable candidate for dynamical age methods. 

\begin{figure*}[h]
    \centering
    \includegraphics[width=0.95\linewidth]{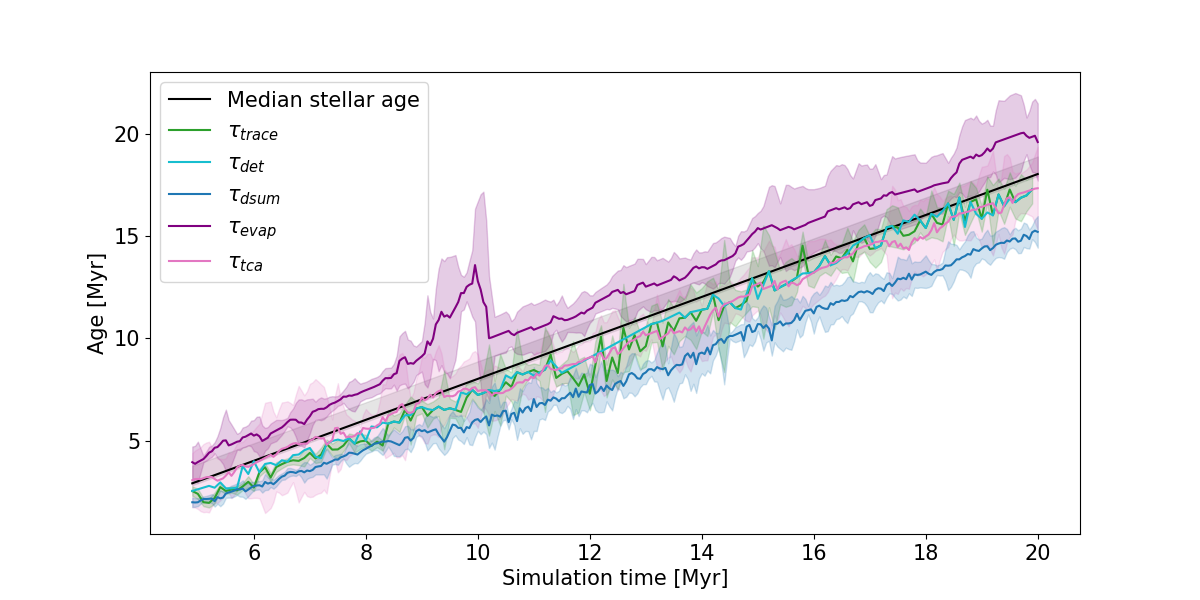}
    \caption{Evolution of different age metrics for $c_2$ as a function of simulation time. Pink and purple lines plot the star-by-star metrics. The blue, green, and teal lines plot the size metric ages. The solid black line plots the median stellar age, with the grey band marking the age of the youngest and oldest star. Typical errors computed using bootstrap error estimation are shown as colored bands around the lines, with a mean of 1 Myr and a maximum of 5 Myr around the spike in $\tau_{\text{evap}}$ at $t_{sim} = 10$ Myr.}
    \label{fig:ages_c2}
\end{figure*}

Fig. \ref{fig:ages_c2} plots dynamical age versus simulation time for $c_2$. We find that size methods can trace the increasing age of $c_2$, with slopes of $1.03\pm 0.01$, $0.98\pm 0.01$, and $0.87\pm 0.01$ for $\tau_{\text{trace}}$, $\tau_{\text{det}}$ and $\tau_{\text{dsum}}$ respectively. The stars-by-star methods have slopes of $1.01 \pm 0.01$ and $1.04\pm 0.01$ for $\tau_{\text{evap}}$ and $\tau_{\text{tca}}$. We note a spike in $\tau_{\text{evap}}$ around $t_{sim} = 10$ Myr. When looking at the cluster during this period, we find that there are two main groupings of stars moving away from the center for the cluster. Because $\tau_{\text{evap}}$ derives an age from the oldest subset of evaporated stars, only 12-14 stars out of the 92 cluster members are used to compute the final age. Therefore, $\tau_{\text{evap}}$ is more sensitive to the presence of substructure than $\tau_{\text{tca}}$.

In both cases, we find that $\tau_{\text{evap}}$ provides an upper limit on stellar ages, and is always larger than $\tau_{\text{tca}}$, while $\tau_{\text{tca}}$ does the best at tracing the isochronal age. For the unbound cluster, size methods are always younger than the median stellar age, with $\tau_{\text{tca}}$ evolving similarly to $\tau_{\text{trace}}$ and $\tau_{\text{det}}$. $\tau_{\text{dsum}}$ consistently produces the youngest dynamical age. This is due to the fact that the robust covariance approach used to determine the minimum size for $\tau_{\text{trace}}$ and $\tau_{\text{det}}$ accounts for velocity outliers and discards them from the overall size determination. Furthermore, the regime for which it is appropriate to use $\tau_{\text{dsum}}$ begins after star formation has finished. This limitation is not present for $\tau_{\text{trace}}$ and $\tau_{\text{det}}$.

\subsection{Size evolution}\label{sec:results-size-evolution}

Methods which estimate the dynamical age from a cluster size assume that the cluster was in a more compact configuration sometime in the past. To investigate whether this assumption holds, we plot the size of $c_2$ as measured by the trace size, $S_{TCM}$, and the determinant size, $S_{DCM}$, for each simulation snapshot. We do not plot the 3D sum of distances because the summation depends on the number of stars which have formed and this changed up to 4.45 Myr. 

\begin{figure*}[h]
     \centering
     \begin{subfigure}[b]{0.48\textwidth}
         \centering
         \includegraphics[width=\textwidth]{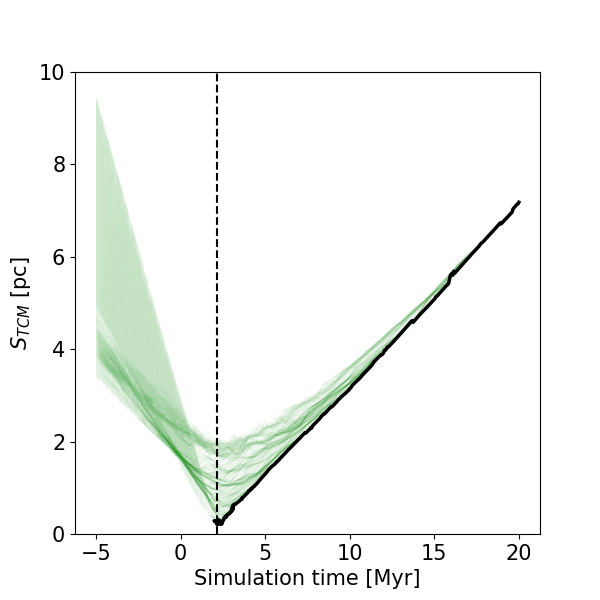}
     \end{subfigure}
     \hfill
     \begin{subfigure}[b]{0.48\textwidth}
        \centering
        \includegraphics[width=\textwidth]{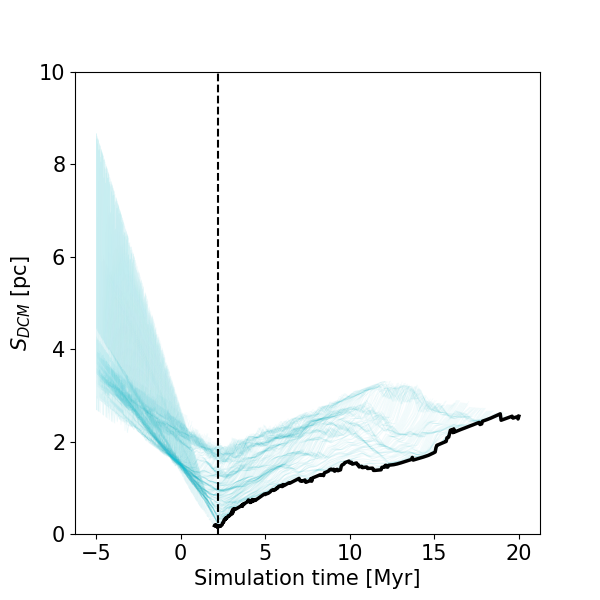}
    \end{subfigure}
    \hfill
    \begin{subfigure}[b]{0.48\textwidth}
         \centering
         \includegraphics[width=\textwidth]{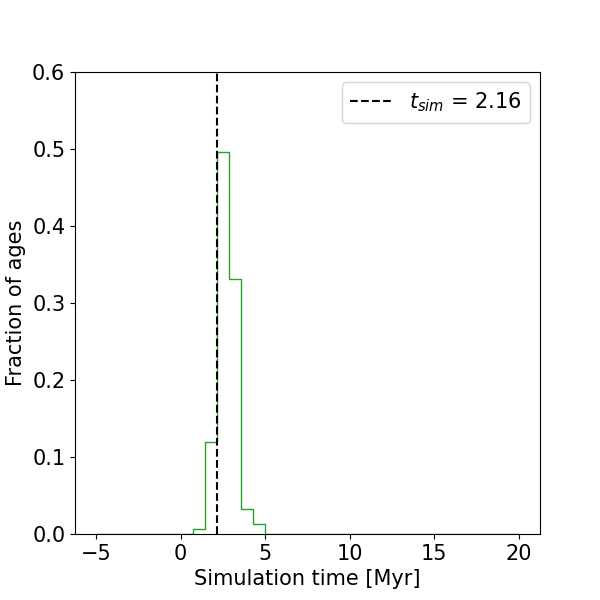}
     \end{subfigure}
    \hfill
    \begin{subfigure}[b]{0.48\textwidth}
        \centering
        \includegraphics[width=\textwidth]{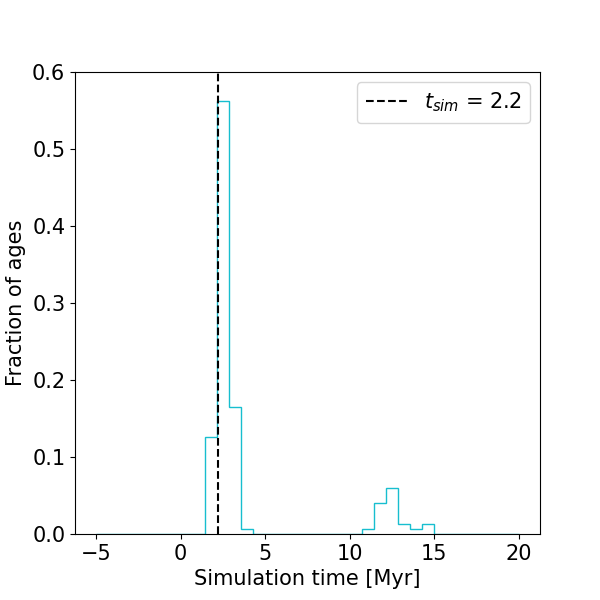}
    \end{subfigure}
    \caption{The evolution of $S_{TCM}$ and $S_{DCM}$ as a function of simulation time. Top: The size evolution of $S_{TCM}$ and $S_{DCM}$ as a function of simulation time (black) plotted on top of the estimated size minimums from the traceback analysis (green and cyan). The black vertical line marks the minimum size in the simulation. Bottom: histogram of the estimated times at which the minimum value of $S_{TCM}$ and $S_{DCM}$ occurred.}
    \label{fig:c2_trace_size}
\end{figure*}

The top row of fig. \ref{fig:c2_trace_size} plots $S_{TCM}$ and $S_{DCM}$ as a function of simulation time. The thick black line plots the size of the cluster in each simulation snapshot while the faint colored curves plot the estimated size of $c_2$ computed using the constant velocity assumption. 

$S_{TCM}$ is minimized at $t_{sim} = 2.16$ Myr and $S_{DCM}$ is minimized at $t_{sim} = 2.2$ Myr. Therefore, the assumption that the cluster was more compact sometime in the past holds. The distribution of $S_{TCM}$ in the bottom left of fig. \ref{fig:c2_trace_size} has a mean of $t_{sim}$ = 2.75 Myr and a median of $t_{sim} = 2.72$, aligning more closely with the onset of expansion in the simulation, occurring at $t_{sim} = 2.76$ Myr, which is 0.60 Myr after the true minimum in the simulation. The distribution in the bottom right of fig. \ref{fig:c2_trace_size} has a mean of $t_{sim}$ = 3.94 Myr and a median of $t_{sim} = 2.63$. 

$S_{TCM}$ uses the trace of the spatial covariance, which can be interpreted as the sum of the spatial variances whereas $S_{DCM}$ uses the determinant, which approximates the product of the variances. This results in $S_{TCM}$ being sensitive to any single variance becoming large, as it adds contributions linearly. $S_{DCM}$ is sensitive to contraction in any dimension, becoming small even when the configuration is more disk-like, and is more sensitive to sample variance and outliers \citep{Sawala2026}. The differences seen fig. \ref{fig:c2_trace_size} come from these different geometrical definitions. 

Size estimators like $S_{TCM}$ and $S_{DCM}$ cannot probe the regime before the onset of cluster expansion, resulting in dynamical ages which align more closely with the onset of expansion than the true minimum size of the cluster. This suggests that size metric methods like $\tau_{\text{trace}}$ and $\tau_{\text{det}}$ do not trace the time at which the cluster's size is minimized. Instead, for unbound systems these methods better determine the onset of cluster expansion. 

\subsection{Velocities}\label{sec:results-velocities}

\begin{figure}
    \centering
    \includegraphics[width=0.95\linewidth]{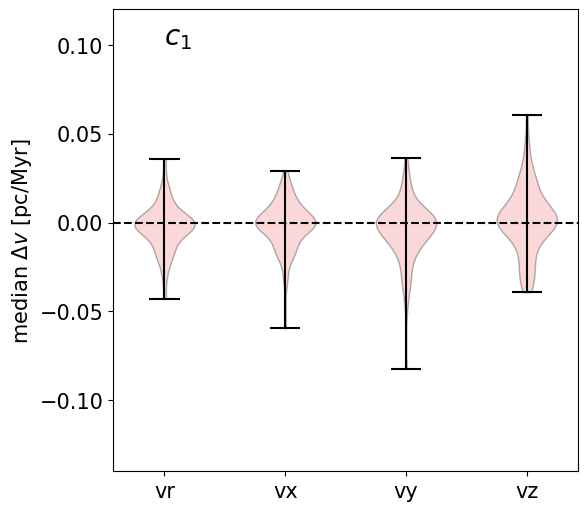}
    \includegraphics[width=0.95\linewidth]{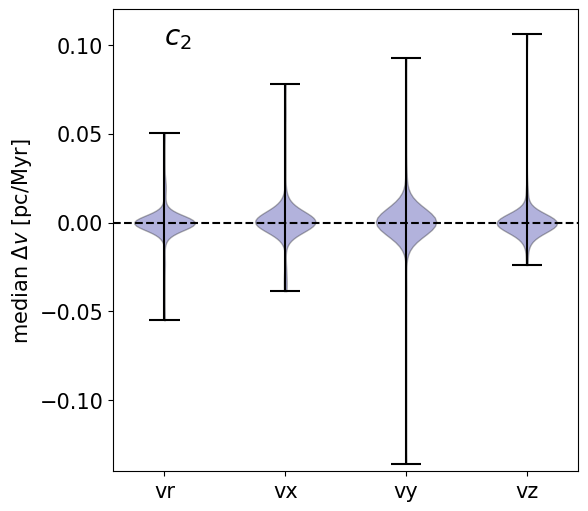}
    \caption{Median change in velocity between snapshots for stars in $c_1$ and $c_2$. Violins plots the 10\%- 90\% range. Error bars plot the minimum and maximum values. For most stars, this median change is small.}
    \label{fig:delta_v}
\end{figure}

All methods explored in this work assume that the velocities of the stars are constant with time. We find that this assumption is not always true, even in unbound systems like $c_2$. To quantify this, we record the median change in velocity between simulation snapshots, spaced 0.05 Myr apart, in the three cartesian coordinates and radially away from the clusters' center, for the stars in $c_1$ and $c_2$. Fig. \ref{fig:delta_v} plots the distribution of median change in stellar velocity for $c_1$ (top) and $c_2$ (bottom). We find that for both $c_1$ and $c_2$, the median changes are centered around zero, with the most extended distribution occurring in $c_2$. For $c_1$, the 10\%- 90\% range is more extended than for $c_2$, meaning that the velocity of the stars in $c_1$ typically change more than the stars in $c_2$, consistent with fact that $c_1$ is mostly bound, with stars orbiting the center of the cluster. 

Despite the fact that the constant velocity assumption does not strictly hold, the dynamical ages discussed in this work produce sensible trends. Star-by-star methods and the robust covariance matrix used to compute $\tau_{\text{trace}}$ and $\tau_{\text{det}}$ remove stars whose velocities do not align with that of the overall cluster, remove stars that are orbiting the center of the cluster and preferentially select stars on radial trajectories. Star-by-star methods like $\tau_{\text{evap}}$ and $\tau_{\text{tca}}$ require stars to trace back within either the median stellar distance or the half-mass radius of the cluster, biasing selection towards unbound stars moving radially away from the center of the cluster. For $c_2$, $\tau_{\text{trace}}$ and $\tau_{\text{det}}$ use the robust covariance. This procedure removes stars that are spatial outliers when calculating the size of the cluster. Stars traveling towards to the center of the cluster, and stars that are orbiting the cluster are removed from the covariance calculation, keeping the stars whose velocities follow the overall motion of the cluster. 

So long as a portion of the stars have relatively constant velocity, dynamical ages can reliably measure the cluster age. Both $c_1$ and $c_2$ contain enough stars for the star-by-star methods to successfully calculate dynamical ages. The smallest cluster for which \citet{Pelkonen2024} calculate $\tau_{\text{evap}}$ is Tuc-Hor, with $21$ members, producing 8 evaporated stars of which 3 are used to derive the final evaporation age. Provided the cluster has few divergent stars, dynamical ages can be computed for small clusters like Tuc-Hor.

In the case of $c_2$, the robust covariance is able to successfully find the spatial covariance of the stars at each time step. This means that dynamical ages methods can use a constant velocity assumption provided that the method contains steps to remove stars with velocities inconsistent with the overall cluster motion. 

\section{Summary and Discussion} \label{sec:discussion}

\subsection{Summary}
In this work we applied five dynamical age methods to simulated star clusters to investigate which processes are traced by different dynamical age methods in a controlled way. These ages are categorized as either size metrics, which derive a single number to describe the spatial extent of the cluster, or star-by-star methods which derive an age from the distribution of times at which the distance between individual stars and the clusters' center is minimized. We analyze one bound cluster and one unbound cluster identified 20 Myr into the simulation. We find that different dynamical age methods trace different physical processes in the formation and assembly of star clusters. Stars-by-star methods produce dynamical ages which increase with simulation time for both the bound and unbound systems, with $\tau_{\text{evap}}$ providing an upper limit for stellar ages and $\tau_{\text{tca}}$ tracing the median age of the stars for both $c_1$ and $c_2$. Size metric methods are only successful when applied to unbound systems whose stars are moving radially away from the cluster's center, a result which is consistent with interpretations from the literature \citep[see][]{Miret-Roig2020, Kerr2022}. $\tau_{\text{trace}}$ and $\tau_{\text{det}}$ appear to trace the onset of cluster expansion, which is not necessarily the same as the time at which the cluster was most compact. In the case of $c_2$ we find that $\tau_{\text{tca}}$ evolves similarly to $\tau_{\text{trace}}$ and $\tau_{\text{det}}$, producing ages which are consistent with each other despite differing approaches in the age derivation. Finally, $\tau_{\text{dsum}}$ produces the youngest dynamical age estimate, consistently younger than the youngest stars within $c_2$. We emphasize that all stars in this simulation formed from the same cloud. Therefore, it is certain that our results are not impacted by stellar contaminants (stars which formed from a different star formation event but is spatially coherent with the cluster of interest). Our work highlights the importance of choosing an appropriate dynamical age method for a given science question.

\subsection{Dynamical ages in relation to cluster formation}

When attempting to analyze dynamical ages, it is important to think about the stages in a star cluster's life. The main points in a cluster's assembly can be summarized as the onset of star formation, the peak of star formation, the end of star formation, and gas expulsion. With the advent of the \textit{Gaia} space telescope, the minimum spatial extent of the cluster and the onset of cluster expansion have become additional phases of the assembly process receiving attention in the literature. In fact, the idea that a cluster experienced some minimum size motivated the development of size metric dynamical traceback ages. 

In our simulation, both the minimum cluster size and time at which the cluster begins expanding, occur before gas expulsion, despite what has been previously assumed \citep{Miret-Roig2024}. Furthermore, we find that star formation continues after the onset of expansion, indicating that gas expulsion is a continuous process, not an instantaneous one. Global expansion begins at $t_{sim} = 2.67$ Myr. In the case of $c_1$, stars continue forming after the onset of expansion. For example, $c_1$ contains stars which formed up to $t_{sim} = 3.42$ Myr, 0.75 Myr after the onset of expansion. 

While no method perfectly traces a specific phase of a star cluster's formation in all cases, the relative evolution between each method directly reflects the evolution of the cluster. \citet{Pelkonen2024} constructed the evaporation age method to trace the onset of star formation. By selecting stars with the greatest individual evaporation ages, $\tau_{\text{evap}}$ is biased towards larger ages when compared to the other star-by-star age, $\tau_{\text{tca}}$. Stars may form in the outskirts of the cluster, some distance from the cluster's center, but trace back to a closer distance. In all cases, we find that $\tau_{\text{evap}}$ provides an upper limit on the median stellar age of the cluster. We find that $\tau_{\text{tca}}$ does the best job of tracing the median stellar age of all the clusters investigated in this work, both bound and unbound. Our work suggests that $\tau_{\text{tca}}$ is the best proxy for isochronal age, and it can be applied to any system displaying signs of expansion, regardless of whether the system is bound or unbound. 

Previous works have shown that properties of cluster formation such as the gas expulsion timescale \citep[e.g.][]{Farias2026}, initial virial parameter of the giant molecular cloud \citep[e.g.][]{Claude2023}, early formation of massive stars \citep[e.g.][]{Lewis2023}, and merger history \citep[e.g.][]{Karam2025} 
can change the global kinematic properties of clusters. However, we do not expect the dynamical age methods to trace different processes than established in this work. For example, the evaporation age method will still select the stars with the greatest individual evaporation ages, and the size metrics will still trace the onset of expansion in unbound clusters.

For star-by-star methods to accurately trace the trajectories of individual stars, the analysis must be performed in the cluster's frame of reference and the velocities of individual stars be taken with respect to the motion of the cluster as a whole. This requires that a cluster center and velocity be defined. Failing to do so causes star-by-star methods to compare the trajectories of the individual stars to the present day position of the cluster's center, instead of the position the cluster's center would have had in the past. $\tau_{\text{tca}}$ is derived using the cluster's center of mass position and velocity whereas $\tau_{\text{evap}}$ is derived using the mean position and velocity of the cluster. In comparing the two approaches, we find that both methods yield centers and cluster velocities that are close to each other, and have minimal impact on the dynamical age methods. The 3D sum of distances also requires a center to be defined. However, in this case, the relative motion between the stars and the cluster center is less important because the method uses all the member stars to define its size. In the case of a simulation, finding the center of mass of the system is simple. Observationally, defining a cluster center and motion can be challenging in systems of few stars, or systems with large astrometric uncertainties.

All the size metrics are sensitive to kinematic outliers \citep{Miret-Roig2020, Quintana2022,Couture2023}. Since the velocities of the stars do change with time, deciding which stars to include in analysis can become difficult. In a simulation, it is known that every star formed from the same cloud. We found that the robust covariance estimator treated kinematic outliers in a systematic way, giving consistent age results without requiring us to discard stars which seemed to be kinematic outliers but were actually a part of the cluster. Had we removed every star which seemingly deviated from the trajectory of the cluster as a whole, we would have removed stars which were real cluster members. We caution against conflating kinematic outliers as contaminants to the system. 

In this work, we know that all the stars originated from the same cloud. Stars which may appear to be kinematic outliers, tracing back to a different point in space from the rest of the cluster, formed alongside the rest of the stars. Furthermore, we know from the simulation that both clusters investigated in this work did have a more compact configuration at early times. This means that we are able to directly test the constant velocity assumption without worrying about contaminants. This is not possible in observations, which must derive algorithms to remove potential contaminants before computing dynamical ages \citep{Couture2023}. Calculating dynamical ages for a simulated cluster allows us to directly study how dynamical ages evolve because the evolution the cluster is known.  

Observationally, cluster membership selection is one of the greatest challenges when performing dynamical age and traceback analysis. Observers will use a combination of spatial and velocity cuts, as well as colour cuts to arrive at the final membership lists. Spatial clustering tends to be done using algorithms such as DBSCAN, HDBSCAN, or Gaussian mixture models \citep[e.g.][]{Miret-Roig2020,Squicciarini2021, Quintana2022, Kerr2022,Couture2023, Kerr2023, armstrong2024,Kerr2025, armstrong2025}. 

After the spatial clustering is performed, observers perform different membership cuts to determine which stars should be included in the final calculation. One approach is to remove stars which trace outside a 3$\sigma$ threshold in any spatial direction at any time step. This is repeated iteratively until no more stars are removed \citep[e.g.][]{Couture2023}. While cuts like this are needed to ensure that stellar contaminants are removed, stringent cuts on stellar traceback risks removing bona fide cluster members whose velocities have been influenced by dynamical interactions or substructure within the cluster. The nuances of this process can directly effect the result of the dynamical age calculation. For example, \cite{Miret-Roig2020} compute $\tau_{\text{trace}}$ and $\tau_{\text{det}}$ for the $\beta$ Pictoris moving group, measuring a dynamical traceback age of $18.5^{+2.0}_{-2.4}$ Myr using a sample of 28 stars. This analysis is repeated by \cite{Couture2023}, who employ a different membership algorithm and find a dynamical age of $20.4 \pm 2.5$ Myr using 25 stars. This result is almost $2$ Myr older than the age published by \cite{Miret-Roig2020} but still $1-3$ Myr younger than the ages obtained by isochrone and lithium depletion boundary methods. Because dynamical ages, particularly size metric methods, are so sensitive to kinematic outliers the choice to include or exclude even a single star can substantially alter the final age. Our investigation shows that stellar velocities are not constant, and so discarding stars based purely on a linear traceback can result in the removal of genuine member stars.

In this idealized set up, we do not consider observational uncertainties nor do we place our cluster within a Galactic potential. This is done to ensure direct comparison between different dynamical age methods. Previous works have shown that the introduction of \textit{Gaia}-like uncertainties result in variation of the estimated dynamical traceback ages of $\sim 2.5$ Myr for size metric methods \citep{Farias2025}. The typical errors on the size metric methods applied to $c_2$ was about 1 Myr. \cite{Pelkonen2024} found that for realistic observational errors on position and velocity, evaporation ages have a $\sim 30\%$ uncertainty of clusters of $\sim 70$ stars, decreasing to $\sim 10\%$ for clusters of $\sim 150$ stars or more.

\citet{Miret-Roig2020} explored the effects of different Galactic potentials on the dynamical age of the $\beta$ Pictoris moving group, investigating several axisymmetric \citep{Allen1991,bovy2015, McMillan2017} and non-axisymmetric models accounting for spiral arms \citep{Drimmel2001, Pichardo2003}. They found that variations in $\tau_{\text{trace}}$ and $\tau_{\text{det}}$ were minimal for their sample, citing short integration times as the reason for minimal dependency on the choice of Galactic potential. Variations in the dynamical ages produced by different Galactic potentials was lower than their membership uncertainties. Therefore they concluded that choice in potential model has minimal impact on the overall dynamical age of a system. 

\section*{Acknowledgments}
A.S is supported by NSERC. C.C.-C. is supported by a postdoctoral fellowship from NSERC. This research was enabled in part by support provided by the Digital Research Alliance of Canada (\url{https://www.alliancecan.ca/en}). 

\section*{Data Availability}
The simulations underpinning this work can be provided upon reasonable request to the corresponding author.

\section*{Software}

Python packages used for data analysis and visualization include \texttt{matplotlib} \citep{Matplotlib2023}, \texttt{numpy} \citep{Harris2020}, \texttt{scikit-learn} \citep{Pedregosa2011}, \texttt{scipy} \citep{2020SciPy-NMeth}, and \texttt{yt}~\citep{yt}.

\bibliographystyle{aasjournalv7.1}
\bibliography{bib}

\end{document}